# Prediction of Students' performance with Artificial Neural Network using Demographic Traits

Adeniyi Jide Kehinde[1], Abidemi Emmanuel Adeniyi[1], Roseline Oluwaseun Ogundokun[1[1111-2222-3333-4444]], Himanshu Gupta[2], Sanjay Misra[3]

[1]Department of Computer Science, Landmark University Omu Aran, Nigeria
[2]Birla Institute of Technology Pilani, Hyderabad
[3]Department of Electrical and Information Engineering, Covenant University, Ota, Nigeria

Adeniyi.jide@lmu.edu.ng, adeniyi.emmanuel@lmu.edu.ng, ogundokun.roseline@lmu.edu.ng, f20150339h@alumni.bits-pilani.ac.in sanjay.misra@covenantuniversity.edu.ng,

**Abstract.** Many researchers have studied student academic performance in supervised and unsupervised learning using numerous data mining techniques. Neural networks often need a greater collection of observations to achieve enough predictive ability. Due to the increase in the rate of poor graduates, it is necessary to design a system that helps to reduce this menace as well as reduce the incidence of students having to repeat due to poor performance or having to drop out of school altogether in the middle of the pursuit of their career. It is therefore necessary to study each one as well as their advantages and disadvantages, so as to determine which is more efficient in and in what case one should be preferred over the other. The study aims to develop a system to predict student performance with Artificial Neutral Network using the student demographic traits so as to assist the university in selecting candidates (students) with a high prediction of success for admission using previous academic records of students granted admissions which will eventually lead to quality graduates of the institution. The model was developed based on certain selected variables as the input. It achieved an accuracy of over 92.3 percent, showing Artificial Neural Network potential effectiveness as a predictive tool and a selection criterion for candidates seeking admission to a university.

**Keywords:** Admission, Student, Quality education, demographic traits, ANN

## 1 Introduction

The scholastic accomplishment of students in the university is the most significant criterion to determine the nature of university students [1-3]. Studying the academic achievements of students is therefore of great importance for fostering student improvements and enhancing the standard of higher education [4-6]. However, this accomplishment is influenced by demographic factors, psychometric factors and previous results of the student [7]. Usually, most higher institutions use the previous academic record of a candidate for admission [8]. While this is okay, the previous result is still influence to a degree by certain demographic factors. The quality of admitted students has a big influence on the institution's level of academic performance, research, and training. The principal goal of the admission framework is to decide the applicants who might perform well in the wake of being admitted into the school. Failure to make an accurate decision on admission can result in the admission of unsuitable student to the institution. Hence, it is essential to analyse the academic potential of students [9, 10].

Majority of the prevalent studies have usually relied majorly on the academic factors for performance prediction [11-14]. While this might be true for a student already admitted into the university, prospective candidates only have previous academic level result. In order to circumvent this setback, there is need to examine the influence of demographic factors on academic performance (since it influences a student's academic performance). This study approaches student's performance prediction with the use of a machine learning technique. Machine learning is mostly used to study complex relationships between data [9, 15]. Machine learning has the ability to learn without necessarily being pre-programmed. Among the machine learning techniques is Artificial Neural Network [16]. Artificial Neural Network (ANN) is growing and gaining recognition in data analysis. It is capable of analysing complex data sets as well as identify relationships between variables [17, 18].

The aim of this study is to implement a neural network which predicts student performance with respect to demographic factors (arguments) such as: gender, age, family background and so on. This is aimed at reducing the incidence of admitting poor student. It also points out students who qualified for admission, but are likely to require attention to avoid failure [19]. These factors have been carefully studied and coordinated to be suitable for computer coding using the Artificial Neural Network model. The system was trained and tested using data of students who have already graduated with a view to enhance predictive device accuracy.

The remaining sections of this paper are organized as follows: section two examined recent and relevant literatures to the study, section three expatiate the methods used in the study and the fourth section shows the result obtained from the proposed methods.

## 2 Literature Review

Some studies have examined student performance prediction, though it has been using previous educational result to predict future result in the same educational level. Some of these studies are reviewed below.

In Vairachilai and Vamshidharreddy [18] a comparison between several classification algorithms which include decision tree, support vector machine (SVM) and Naïve Bayes was presented. Naïve Bayes performed best with an accuracy of 77%. Li, Zhu, Zhu, Ji and Tang [20] used the internet for student performance prediction. Online learning records for project course and network logs were used in the prediction. Spatial Prediction based on Deep Network (SPDN) was used for predicting student performance at a course. They reported an accuracy of 73.51%. An approach to boost student performance prediction using interactive online pool and further considering student features interaction was proposed by Wei, Li, Xia, Wang and Qu [21]. New features such as time, first attempt and first drag and drop were introduced. Yahaya, Yaakub, Abidin, Razak, Hasbullah and Zolkipli [22] they attempted to predict the performance of undergraduate students in chemistry courses using a multilayer perceptron. An accuracy of 92% was recorded. The use of ensemble model for the improvement of student graduation prediction was studied by Lagman, Alfonso, Goh, Lalata, Magcuyao and Vicente [23]. This is aimed at identifying students who have a high chance of not meeting the graduation time. With the identification proposed, such students can be given attention in areas that they are deficient. An average accuracy of 88.30% was recorded at best.

Quinn and Gray [24] approached student's academic performance prediction with Moodle data using a further education setting. Their study investigated if data from the learning management system Moodle can be used in predicting student's academic performance. This is aimed at predict-ing whether a student would pass a course or not. The classifier on all course data had an accuracy of 65%. Predicting whether a student would pass or fail very well has an accuracy of 92.2%. Data built on the first six weeks performed poorly and there was a need to extend the data gathered. Li, Wei, Wang, Song and Qu [20] proposed the use of Graph Neural Network (GNN) for better student performance prediction. Interactive online questions were presented to students. A new GNN was further presented, that is R2- GNN. Rakic, Tasic and Marjanovic [25] presented a paper aimed at examining how useful and impactful the digital technology is on an e-learning platform. Data from online sources were used to examine key pointers of the performance of students in several courses. Social Network Analysis, K-Means clustering and Multiple Linear Regression were used to evaluate student's success. The result showed a huge relationship between the performance of the student and the use of digital educational resources from the e-learning platform. There are several other related studies[26-27] available in literature related to students performance which we are not considering due to limit of the work.

## 3 Material and Method

It has been observed that a number of variables (including demographics, extra-curricular activities, environment as well as biological factors) are considered to affect the performance of a student. These variables were carefully researched and modified to form a detailed equivalent number appropriate for computer coding. These variables are taken as input to the system. The block diagram of the system is depicted in figure 1.

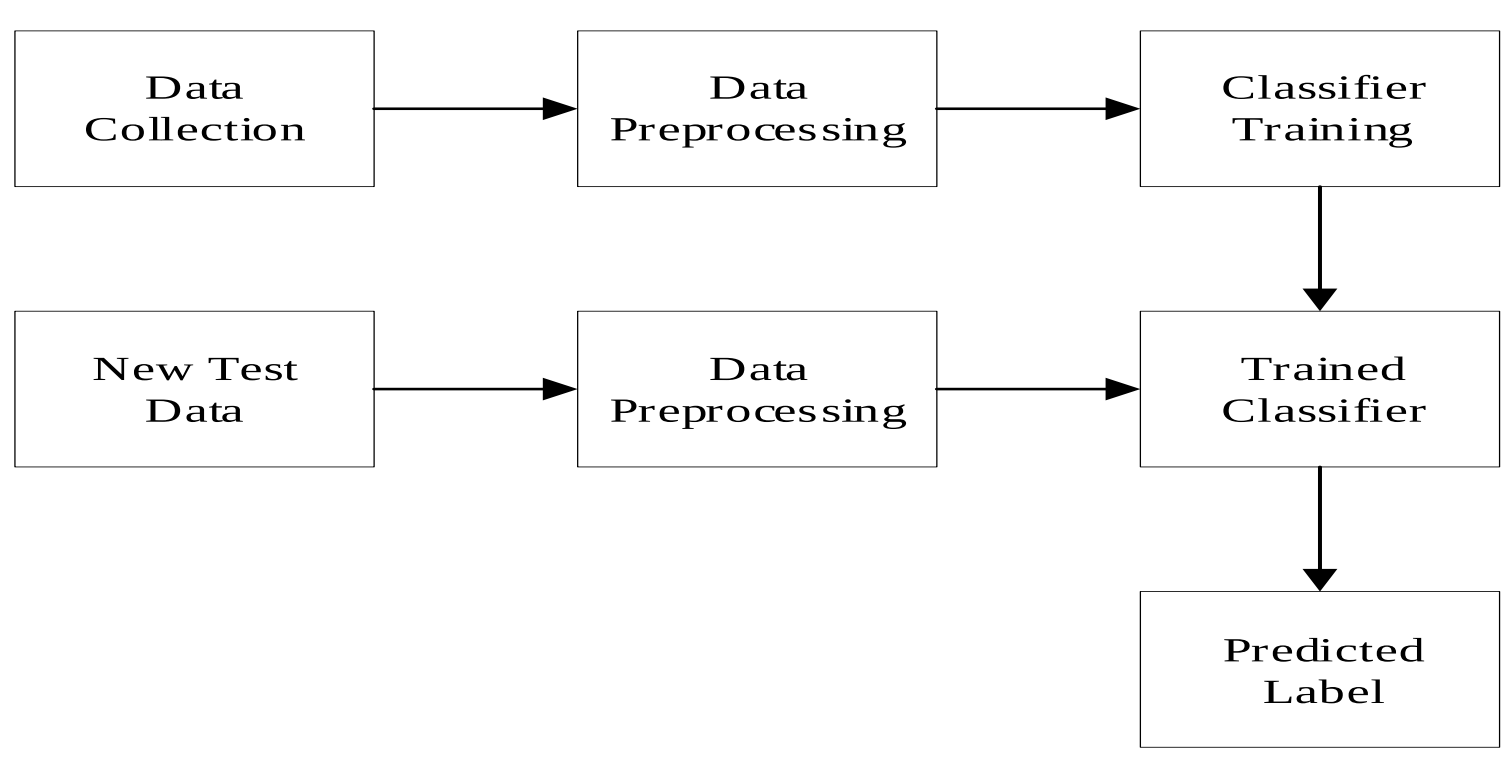

**Fig. 1.** Block diagram of the proposed system

### 3.1 Data Collection

The input variables used are those gotten from the dataset of the student database. The dataset was obtained from the UCL Machine Learning Repository (https://www.kaggle.com/dipam7/student - grade-prediction). The dataset includes a number of variables, among which using educational data mining the features that are most related to and have more effect on the output variables were selected and they include: The output variable reflects a student's success at the end of a term/semester. However, for this study, the variables used include G1(assignments), G2(tests), G3(Final exam) which are numeric values within the range of 0 to 20.

Since the classifier accepts input in numeric form, there is need to converting categorical variables to numerical in preparation for training. Table 1 shows the transformed data.

Figure 2, Figure 3 and Figure 4 below is a probability plot of sample data against the quantiles of a given theoretical distribution (the default normal distribution). It basically calculates optionally a best-fit line for the data and plots the results. From Fig 1, the skewness was determined to be 0.240613 while the Kurtosis was -0.693830. The mean, mu = 10.91 and the standard deviation, sigma = 3.31. The skewness in the probability plot of G2 in Fig 2 was determined to be -0.431645 while the Kurtosis was 0.627706. Finally, the skewness was determined to be -0.431645 while the Kurtosis was 0.627706 for G3 in Fig 3. This skewness and kurtosis values for G1 implies that the distribution is fairly symmetrical and lack outliers. However, the value of skewness and kurtosis for G2 and G3 shows that they a slightly negative skewed and has outliers. Hence, the correlation matrix was used to eliminate the outliers in the distribution.

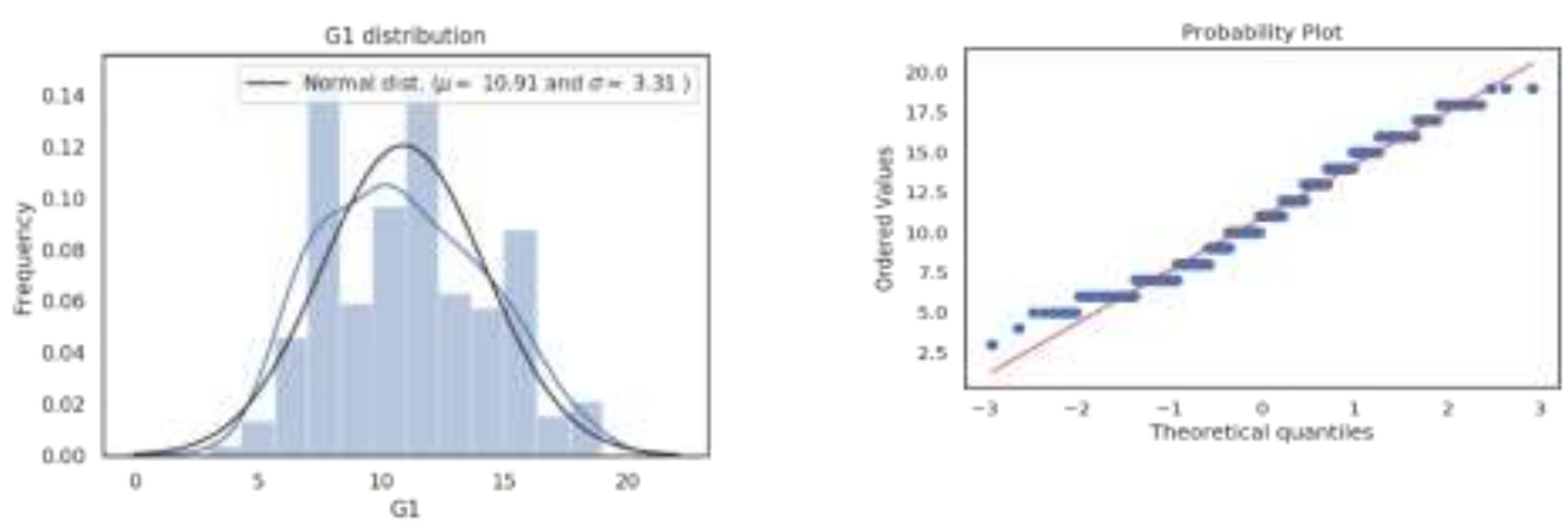

**Fig. 2.** Probability plot of G1

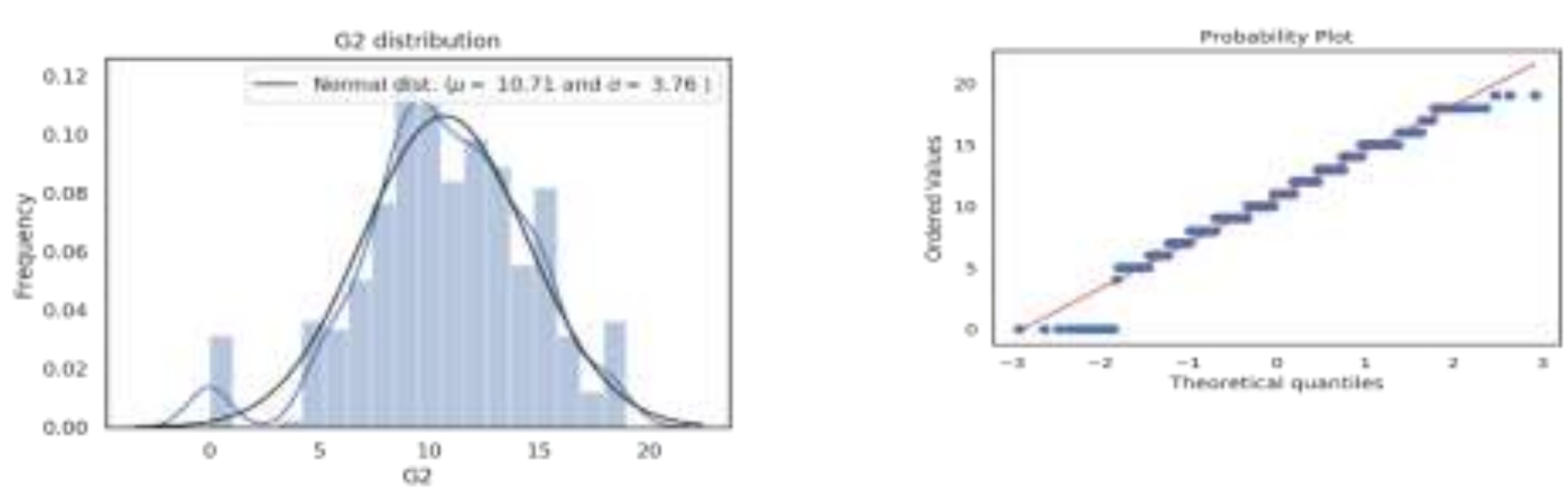

**Fig. 3.** Probability plot of G2

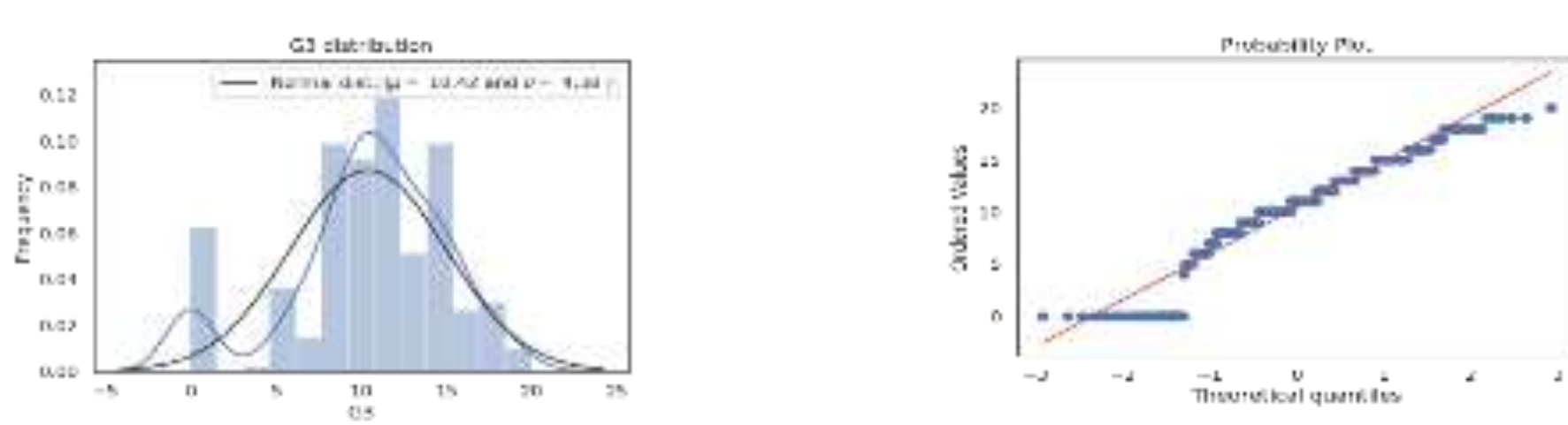

**Fig. 4.** Probability plot of G3

**Features Selection using the Correlation Matrix.**

Correlation examines the relationship and association that exist between variables. The relationship and association refer to how much a variable is affected by a change in another variable. Correlation could be simple, partial or multiple depending on the number of variables being examined at once. A zero correlation typically implies no relationship exists between the variables. The correlation matrix was applied to determine the features that actually affect the result G1, G2 and G3. This weeds out the variables (features) that are not relevant to the performance of the candidate (feature selection) [7]. The correlation matrix in Figure 5 generated through the exploratory data analysis finds which features are most related to the G1, G2, or G3 scores.

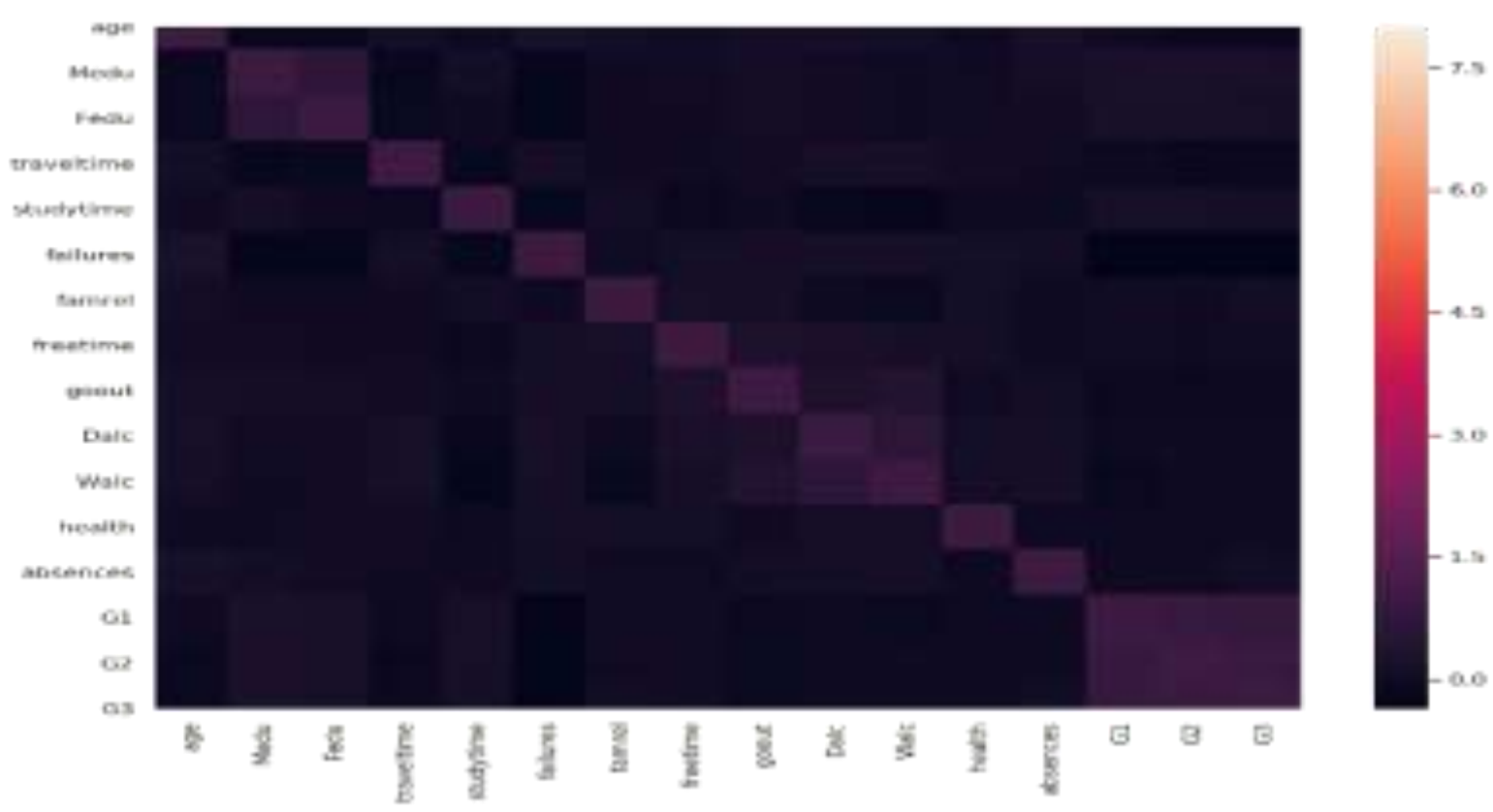

**Fig. 5.** Correlation matrix

Where G1, G2 and G3 stand for assignments scores, test scores and final exam scores respectively. The most correlated features are shown in Table 1.

**Table 1.** Most correlated features

| | |
|---|---|
| 0 | G1 |
| 1 | G2 |
| 2 | G3 |
| 3 | Medu |
| 4 | Fedu |
| 5 | Studytime |
| 6 | famrel |
| 7 | freetime |
| 8 | absences |
| 9 | age |

### 3.2 Prediction

For predicting student's performance, Artificial Neural Network was applied. Artificial Neural Network (ANN) is a system that is made up of a number of neuron (taken as input to the system), hidden layers and output layers [28]. Weights are used to connect the neurons. The process of designing the network is a tedious one as there is dependency between the various variables like the structure of the network and the values of the weights [29, 30, 19]. The learning phase and the prediction phase are the two phases of the Network. The learning phase plays an important role as this is where the weight is adjusted to produce the required output [28]. The adjustment of the weight is repeated until a termination criterion is met [31,32]. This termination criteria could be the acceptable mean square or number of evolutions to achieve the target value. After training an ANN, the trained network is tested and validated. To test and validate the trained ANN, unknown samples are used to test the Network. To express an ANN problem, we can use the set of inputs in (1) (Gupta, Khosravy, Patel, Gupta & Varshney, 2020),

$$X = \{x = x_i | x \in \mathbb{R}^n, i = 1,2,\ldots,n\} \quad (1)$$

and outputs in (2)

$$Y = \{y = y_o | y \in \mathbb{R}^m, o = 1,2,\ldots,m\} \quad (2)$$

From (1) and (2), we can observe that the input and output layers of the ANN have n and m neurons respectively. Given a chosen number of hidden layers, to design an ANN for the data in (1) and (2) the objective function can be given as in (3).

$$\mathfrak{f} = f(X, Y, W) \quad (3)$$

$W$ being the weight connections between the layers. It could be represented as in (4).

$$W = \{W_{12}, W_{23}, \ldots, W_{l-1,l}\} \quad (4)$$

$l$ is the sum of the layers, this include the input, hidden and output layers. Minimizing $\mathfrak{f}$ in (3), the mean square error (MSE) gives an optimized synaptic weight of ANN.

In [33], from schematic diagram of an ANN, it is observed that the system performs addition of all the inputs first, and then the output of the addition is passed to the transfer function. The output of the transfer function serves as an input to the neuron in the preceding layer. Each neuron is formed with the representation in (5) [34].

$$y_p = f\left(\sum_{i=0}^{n} w_i^p x_i^p + b^p\right) \quad (5)$$

$x_i^p$ in (5) is the $ith$ input of the $pth$ neuron, $w_i^p$ is the weight value of the connection between $ith$ input and $pth$ neuron and $\mathfrak{f}$ is the transfer function. A bias is associated with each neuron and it is represented with $b$. $y_p$ is the output of the $pth$ neuron which is passed as input to the next neuron?

## 4 Results and Discussion

This section describes the statistical results, configuration of the Artificial Neural Network and performance based on the academic performance educational data obtained of 376 students. Python programming language was used as well as the TensorFlow backend.

### 4.1 Dataset grouping and Network Architecture

In supervised training, the data set is divided into three groups: the training set, validation or verification test and testing set. The training set allows the system to detect the relationship between the input data and the given outputs, so as to create a relation between the input and the expected outcome. Altogether, 396 student records were used in this analysis where the training and testing set was split in the ratio 70:30. The architecture of the system is shown in table 3.

**Table 2.** Architecture of the chosen network model

| Layer (type) | Output Shape | Param # |
|---|---|---|
| dense_1 (Dense) | (None, 80) | 2480 |
| dropout_1 (Dropout) | (None, 80) | 0 |
| dense_2 (Dense) | (None, 120) | 9720 |
| dropout_2 (Dropout) | (None, 120) | 0 |
| dense_3 (Dense) | (None, 20) | 2420 |
| dropout_3 (Dropout) | (None, 20) | 0 |
| dense_4 (Dense) | (None, 10) | 210 |
| dropout_4 (Dropout) | (None, 10) | 0 |
| dense_5 (Dense) | (None, 3) | 33 |

The network architecture in table 3 shows how the model is structured. It shows all the layers the model consists of as well as the parameters the model consists of. Dense 1 to dense 5 denotes each dense layer and the dropout layers are added after each layer to reduce dimensions. The dimension reduction is essential for feature extraction, to take out the features that are important in the dataset. The params denotes parameters which show the number of available data in the dataset. After every dropout layer, the number of parameters reduces.

The network was trained with batch size set to eight and the Epoch set to terminate at 500. Figure 6 and 7 shows the histogram of the training and validation in terms of its Mean Absolute Error (MAE) and loss.

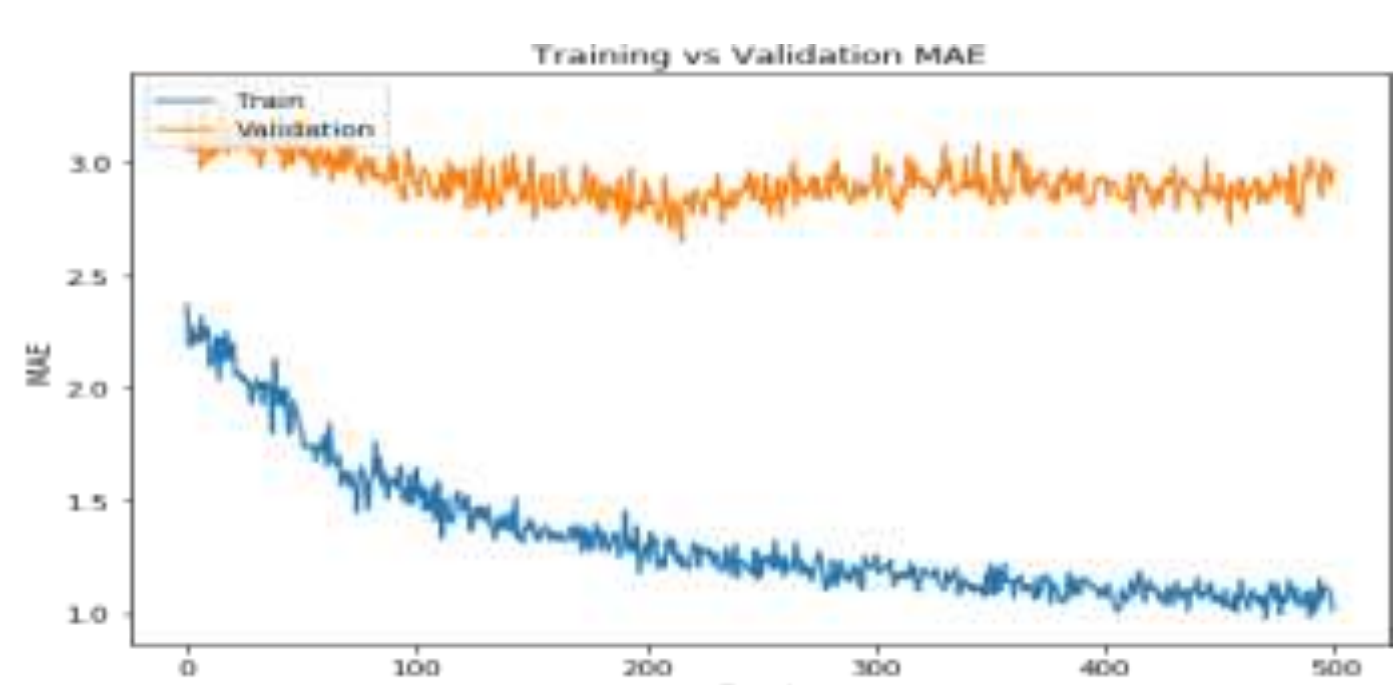

**Fig. 6.** Training and validation MAE

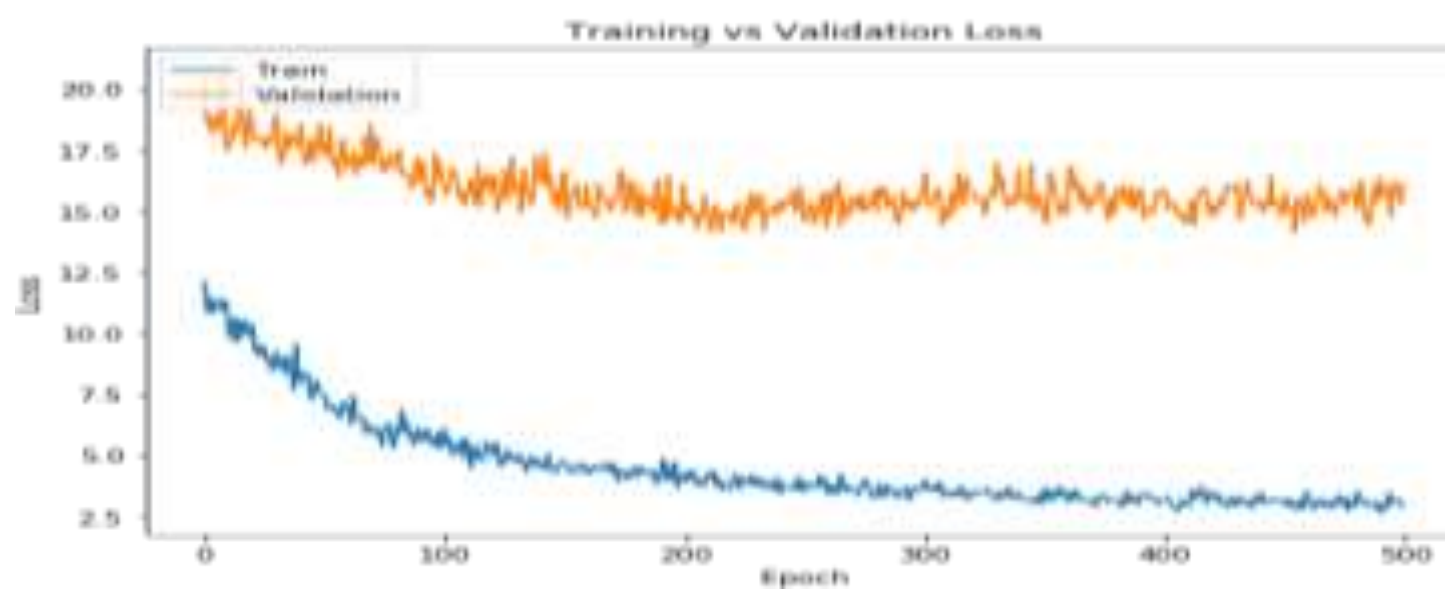

**Fig. 7.** Training and validation loss

### 4.2 Discussion

An accuracy of 92.26% was obtained after testing the system. Table 4 shows a comparison of the result obtained with other similar system.

**Table 3.** Comparison of Similar Student Performance Prediction Systems

| Authors | Study focus | Accuracy (%) |
|---|---|---|
| Vairachilai and Vamshidharreddy [18] | Comparison of methods – decision tree, SVM and Naïve bayes. | 77 |
| Li, Zhu, Zhu, Ji and Tang [20] | Performance at a course | 73.51 |
| Yahaya et al. [22] | Undergraduate who will pass chemistry courses | 92 |
| Quinn and Gray [24] | Pass or fail on All courses | 65 |
| Pekuwali [26] | Student performance at the final year. | 94.24 |
| Proposed System | Student will graduate or not | 92.26 |

Considering table 4, it can be noted that the performance of the system is encouraging. In the instance where the accuracy was higher than the proposed system, previous records of academic performance were being used to predict subsequent result. This makes it difficult to use the system to predict student's performance at the time of admission. The designed system considers only the demographic factors and its performance shows a high accuracy. The performance of the system was greatly influenced by the methods used in examining the correlation between the input data and the predicted outcome.

## 5 Conclusion

This paper presented an approach to student performance prediction using demographic features only. The relationship between the demographic features and the result was examined with the skew and kurtosis of the dataset. Outlier from the statistical evaluation of skew and kurtosis were further removed in the feature selection stage using the correlation matrix. The resulting dataset was then used to train an artificial neural network. The demographic dataset was obtained from the UCL Machine Learning Repository. The student performance prediction showed an accuracy of 92.26%. The proposed approach is not restricted to demographic features only. Hence, examining the performance of the system with both demographic and previous records of students for predicting student's performance at the next educational level could be an area of study in the future.

## References

1. Awotunde, J.B., Ogundokun, R.O., Ayo, F.E., Ajamu, G.J., Adeniyi, E.A., Ogundokun, E.O. (2020). Social Media Acceptance and Use Among University Students for Learning Purpose Using UTAUT Model. Advances in Intelligent Systems and Computing, 1050, pp. 91-102
2. Ogundokun, R.O., Adebiyi, M.O., Abikoye, O.C., Oladele, T.O., Lukman, A.F., Adeniyi, A.E., ..., Akande, N.O. (2019). Evaluation of the scholastic performance of students in 12 programs from a private university in the south-west geopolitical zone in Nigeria. F1000Research, 8
3. Soni A., Kumar V., Kaur R. and Hemavathi D. (2018). Predicting Student Performance using Data Mining Techniques. International Journal of Pure and Applied Mathematics, vol. 119, no. 12, pp. 221-227.
4. Omolewa, O. T., Oladele, A. T., Adeyinka, A. A., & Oluwaseun, O. R. (2019). Prediction of Student's Academic Performance using k-Means Clustering and Multiple Linear Regressions. Journal of Engineering and Applied Sciences, 14(22), 8254-8260.
5. Shahiria A. M., Husaina W. and Rashida N. A. (2015). A Review on Predicting Student's Performance using Data Mining Techniques. The Third Information Systems International Conference; Procedia Computer Science vol. 72, pp. 414 – 422. https://doi.org/10.1016/j.procs.2015.12.157.
6. Imran M., Latif S., Mehmood D. and Shah M. S. (2019). Student Academic Performance Prediction using Supervised Learning Techniques. iJET, vol. 14, no. 14.
7. Altabrawee H., Ali O. A. J. and Ajmi S. Q. (2019). Predicting Students' Performance Using Machine Learning Techniques. Journal of University of Babylon, Pure and Applied Sciences, vol. 27, no. 1.
8. Agrawal H. and Mavani H. (2015). Student Performance Prediction using Machine Learning. International Journal of Engineering Research & Technology (IJERT), vol. 4, iss. 03.
9. Sultana J., Rani M. U. and Farquad M. A. H. (2019). Student's Performance Prediction using Deep Learning and Data Mining Methods. International Journal of Recent Technology and Engineering (IJRTE), vol. 8, iss. 1S4.
10. Wakelam E., Jefferies A., Davey N. and Sun Y. (2019). The potential for student performance prediction in small cohorts with minimal available attributes. British Journal of Educational Technology. http://doi.org/10.1111/bjet.12836.
11. Bergin S., Mooney A., Ghent J. and Quille K. (2015). Using Machine Learning Techniques to Predict Introductory Programming Performance. International Journal of Computer Science and Software Engineering (IJCSSE), vol. 4, iss. 12, pp. 323-328.
12. Ebenezer J. R., Venkatesan R., Ramalakshmi K., Johnson J., Glen P. I., Vinod V. (2019). Application of Decision Tree Algorithm for prediction of Student's Academic Performance. International Journal of Innovative Technology and Exploring Engineering (IJITEE), vol. 8, iss. 6S.
13. Zohair L. M. A. (2019). Prediction of Student's performance by modelling small dataset size. Abu Zohair International Journal of Educational Technology in Higher Education, pp. 16:27. https://doi.org/10.1186/s41239-019-0160-3.
14. Buenaño-Fernández D., Gil D. and Luján-Mora S. (2019). Application of Machine Learning in Predicting Performance for Computer Engineering Students: A Case Study. Sustainability, vol. 11, no. 2833. http://doi.org/10.3390/su11102833.
15. Sekeroglu B., Dimililer K. and Tuncal K. (2019). Student Performance Prediction and Classification Using Machine Learning Algorithms. ICEIT 2019, March 2–4, 2019, Cambridge, United Kingdom. http://doi.org/10.1145/3318396.3318419.
16. Rastrollo-Guerrero J. L., Gómez-Pulido J. A. and Durán-Domínguez A. (2020). Analyzing and Predicting Students' Performance by Means of Machine Learning: A Review. Appl. Sci., vol. 10, no. 1042. https://doi.org/10.3390/app10031042.
17. Hamoud A. K. and Humadi A. M. (2019). Student's Sucess Prediction Model Based on Artificial Neural Networks (ANN) and a Combination of Feature Selection Methods. Journal of SouthWest Jiaotong University, vol. 54, no. 3. https://doi.org/10.35741/issn.0258-2724.54.3.25.

18. Vairachilai S. and Vamshidharreddy. (2020). Student's Academic Performance Prediction Using Machine Learning Approach. International Journal of Advanced Science and Technology, Vol. 29, No. 9s, pp. 6731-6737.
19. Ofori F., Maina E., and Gitonga R. (2020). Using Machine Learning Algorithms to Predict Students' Performance and Improve Learning Outcome: A Literature Based Review. Stratford Peer Reviewed Journals and Book Publishing Journal of Information and Technology, vol. 4, iss. 1, pp. 33-55.
20. Li X., Zhu X., Zhu X., Ji Y. and Tang X. (2020). Student Academic Performance Prediction Using Deep Multi-Source Behavior Sequential Network. PAKDD 2020, pp. 567-579. https://doi.org/10.1007/978-3-030-47426-3_44.
21. Wei H., Li H., Xia M., Wang Y., Qu H. (2020). Predicting Student Performance in Interactive Online Question Pools Using Mouse Interaction Features. LAK 2020, March 23 - March 27. https://doi.org/10.1145/3306307.3328180.
22. Yahaya C. A. C., Yaakub C. Y., Abidin A. F. Z., Razak M. F. A., Hasbullah N. F. and Zolkipli M. F. (2020). The prediction of undergraduate student performance in chemistry course using multilayer perceptron. The 6th International Conference on Software Engineering & Computer Systems; IOP Conf. Series: Materials Science and Engineering 769. https://doi.org/10.1088/1757-899X/769/1/012027.
23. Lagman A. C., Alfonso L. P., Goh M. L. I., Lalata J. P., Magcuyao J. P. H., and Vicente H. N. (2020). Classification Algorithm Accuracy Improvement for Student Graduation Prediction Using Ensemble Model. International Journal of Information and Education Technology, Vol. 10. https://doi.org/10. 10.18178/ijiet.2020.10.10.1449.
24. Quinn R. J. and Graya G. (2020). Prediction of student academic performance using Moodle data from a Further Education setting. Irish Journal of Technology Enhanced Learning, vol. 5, Iss. 1.
25. Rakic S., Tasic N. and Marjanovic U. (2020). Student Performance on an E-Learning Platform. Mixed Method Approach. iJET, vol. 15, no. 2.
26. LuisFernandez-Sanz, Josea-Medina, Maite Villalba, Sanjay Misra, A Study on the Key Soft Skills for Successful Participation of Students in Multinational Engineering Education 'International Journal of Engineering Education, , Vol. 33, No. 6(B), pp. 2061–2070, 2017
27. Luis Fernández, José Amelio Medina, María Teresa Villalba de Benito, Sanjay Misra, 'Lessons From Intensive Educational Experiences For Ict Students In Multinational Settings' Technical Gazette, Vol 24, no 4. 2017
28. Choong C. E., Ibrahim S., El-Shafie A. (2020). Artificial Neural Network (ANN) model development for predicting just suspension speed in solid-liquid mixing system, Flow Measurement and Instrumentation. https://doi.org/10.1016/j.flowmeasinst.2019.101689.
29. Jabamony J. and Shanmugavel G. R. (2020). IoT Based Bus Arrival Time Prediction Using Artificial Neural Network (ANN) for Smart Public Transport System (SPTS). International Journal of Intelligent Engineering and Systems, Vol.13, No.1. https://doi.org/ 10.22266/ijies2020.0229.29.
30. Bui D., Nguyen T. N., Ngo T. D. and Nguyen-Xuan H. (2020). An artificial neural network (ANN) expert system enhanced with the electromagnetism-based firefly algorithm (EFA) for predicting the energy consumption in buildings, Energy 190. https://doi.org/10.1016/j.energy.2019.116370.
31. Solanki R. B., Kulkarni H., Singh S., Varde P. V. and Verma A. K. (2020). Artificial Neural Network (ANN)-Based Response Surface Approach for Passive System Reliability Assessment, Reliability and Risk Assessment in Engineering, Lecture Notes in Mechanical Engineering. https://doi.org/10.1007/978-981-15-3746-2_29.
32. Raya A., Halderb T., Jenaa S., Sahooa A., Ghoshb B., Mohantyc S., Mahapatrad N. and Nayaka S. (2020). Application of artificial neural network (ANN) model for prediction and optimization of coronarin D content in Hedychium coronarium, Industrial Crops & Products, 146. https://doi.org/10.1016/j.indcrop.2020.112186.
33. Gupta N., Khosravy M., Patel N., Gupta S. and Varshney G. (2020). Artificial Neural Network Trained by Plant Genetic-Inspired Optimizer, Frontier Applications of Nature Inspired Computation. https://doi.org/10.1007/978-981-15-2133-1_12.
34. Upadhyay N. and Kankar P. K. (2020). Integrated Model and Machine Learning-Based Approach for Diagnosis of Bearing Defects. Reliability and Risk Assessment in Engineering, Lecture Notes in Mechanical Engineering. https://doi.org/10.1007/978-981-15-3746-2_20.